\documentclass[11pt,a4paper]{article}
\usepackage[utf8]{inputenc}
\usepackage[T1]{fontenc}
\usepackage{lmodern}
\usepackage[english]{babel}
\usepackage[margin=2.5cm]{geometry} 
\usepackage{graphicx} 
\usepackage{booktabs} 
\usepackage{natbib} 
\usepackage{amsmath} 
\usepackage{amssymb} 
\usepackage{setspace} 
\usepackage{caption} 
\usepackage[colorlinks=true, citecolor=blue, linkcolor=blue, urlcolor=blue]{hyperref} 
\usepackage{subcaption} 
\usepackage{longtable}
\usepackage{array}
\usepackage{makecell}  
\usepackage{tabularx}
\usepackage{ragged2e}
\usepackage{float}

\setcitestyle{round}

\let\oldfootnote\footnote
\renewcommand{\footnote}{\fontsize{9}{11}\selectfont\oldfootnote}

\def\anonymous{0} 

\title{NIH-MPINet: A Large-Scale Feature-Rich Network Dataset for Mapping the Frontiers of Team Science}
\ifnum\anonymous=1
    \author{Anonymous Authors}
\else
    \author{
        Cuiran Shi \\
        Department of Biostatistics, State University of New York at Buffalo, NY, USA
        \and
        Shuying Han \\
        Harvard T.H. Chan School of Public Health, MA, USA 
        \and
        Shreya Kusumanchi, Mia Zhou, Didong Li\thanks{Corresponding author: didongli@unc.edu} \\
        Department of Biostatistics, University of North Carolina at Chapel Hill, NC, USA 
    }
\fi
\date{} 

\begin{document}

\maketitle

\begin{abstract}
This study presents a large-scale network dataset, NIH-MPINet, curated from NIH RePORTER and PubMed, characterizing collaboration among multiple Principal Investigators (multi-PIs) on NIH R01-equivalent grants from 2006 to 2023. The network characterizes 30,127 PIs as nodes and their collaborations on 86,743 NIH R01-equivalent grants as edges, spanning 888 recipient organizations and supported by 40 NIH Institutes and Centers. We also curated comprehensive metadata, including node-level features such as PI affiliation, alongside edge-level features comprising grant years, titles, and abstracts. Using these data, we constructed a PI collaboration network and identified 19 communities as well as 20 major research topics. Several collaboration communities showed distinct thematic profiles, such as cardiovascular health, cancer immunotherapy, neuroscience, and microbiome research, while genetics and genomics were broadly represented across communities. By incorporating temporal analysis, we observed shifts in research topics and collaboration patterns over time. Topics like healthcare and outcomes research, cognitive health, and Alzheimer's disease have become more prominent in recent years, whereas molecular and cellular biology has seen a relative decline. Overall, this work provides a high-fidelity, feature-rich resource for advancing statistical learning methods and network analysis-based discoveries in the study of long-term biomedical collaboration.
\end{abstract}


\textbf{Keywords:} Multi-PI NIH grants,  research collaboration networks, community detection, topic modeling


\section{Introduction}

The National Institutes of Health (NIH) serves as the primary engine for biomedical and health-related research in the United States, accounting for over 50\% of all federal funding in this sector. With an annual budget of approximately $\$48$ billion, nearly five times that of the National Science Foundation (NSF), the NIH supports over 300,000 researchers at more than 2,500 institutions globally~\citep{nih_budget}. Within this vast funding landscape, the multiple Principal Investigator (multi-PI) mechanism was formally established in 2006 in response to recommendations from the NIH Bioengineering Consortium and the NIH Roadmap Initiative \href{https://grants.nih.gov/grants/guide/notice-files/not-od-07-017.html}{(Establishment of Multiple Principal Investigator Awards for the Support of Team Science Projects)}. As scientific questions grow more complex and interdisciplinary, these multi-PI mechanisms facilitate team science, an approach recognized for its success in accelerating innovation by integrating diverse methodologies that exceed the capacity of a single-investigator lab~\citep{national2015enhancing}. Understanding these collaboration mechanisms is essential for characterizing modern research practices, evaluating the impact of funding policies on scientific productivity, and identifying emerging interdisciplinary frontiers.


Previous research has leveraged NIH grant data to study research content, funding patterns, and collaboration structures from multiple perspectives. For instance, \cite{talley2011database} created a database in which they used text mining to extract latent categories and clusters from NIH grant titles and abstracts, while \cite{malikireddy2017network} analyzed word co-occurrence networks derived from NIH R01 summary statements to examine linguistic patterns and potential biases in reviewer critiques. Other work has focused on institutional funding distributions, such as the  characterization of NIH awards to diagnostic radiology departments~\citep{franceschi2017patterns}. More recently, researchers have incorporated time-aware topic modeling approaches to track changes in research focus over time~\citep{zhang2023turtling} and proposed grant recommendation systems that align NIH funding announcements with researchers’ publication records~\citep{zhu2023novel}. Complementing these studies, researchers have examined scientific collaboration using large-scale co-authorship and co-citation networks derived from publication data to study the evolution of research communities and interests over time \citep{ji2022co}. 

While these studies demonstrate strong interest in using administrative and bibliometric data to study research communities, collaboration, and thematic trends, existing work typically bifurcates into either the textual analysis of grant-related documents or the structural analysis of collaboration patterns inferred from publication networks. Consequently, there remains a critical gap in research that jointly leverages grant-level metadata, investigator-level attributes, and the explicit representations of multi-PI collaboration. Furthermore, because relevant information is often distributed across disconnected platforms, researchers typically rely on custom extraction workflows that entail significant manual and computational overhead. These limitations highlight the need for a large-scale, unified, and feature-rich resource that integrates these elements, including node-level investigator attributes and edge-level grant metadata, to support reproducible network and statistical learning analyses.


To address these gaps, we introduce NIH-MPINet, a large-scale dataset designed to characterize multi-PI collaboration structures. Our curation focuses on NIH R01-equivalent grants, including activity codes such as R01, R35, R37, DP1, DP2, DP5, R56, RF1, RL1, and U01, as these represent the primary mechanisms for high-impact, investigator-initiated research where multi-PI synergy is most influential. While publicly available resources such as \href{https://reporter.nih.gov}{NIH RePORTER} provide extensive information on funded projects, they are primarily designed for administrative reporting rather than large-scale, reproducible network analysis. Critical data points such as PI roles, organizational affiliations, and research abstracts, are often fragmented across platforms, recorded at inconsistent levels of detail, or buried in unstructured formats. Consolidating these records from 2006 to 2023 required a non-trivial pipeline for data cleaning, entity resolution, and temporal alignment. By integrating records from NIH RePORTER and PubMed, NIH-MPINet provides a unified representation of 30,127 PIs (nodes) and the 86,743 grants they served as co-PIs (edges), enriched with node-level features (e.g., PI affiliations) and edge-level metadata (e.g., grant years, titles, and abstracts).

Beyond data curation, we provide a comprehensive analysis framework to demonstrate the dataset’s potential for uncovering the hidden architecture of biomedical collaboration. In particular, we employ the Leiden algorithm to identify 19 distinct communities, reflecting the structural organization of investigator teams. To characterize the intellectual content of these collaborations, we apply BERTopic modeling to extract 20 major research themes from grant titles and abstracts (edge-level features). By mapping these 20 topics onto the 19 communities, we identify specialized collaborative hubs, ranging from neuroscience to cancer immunotherapy, and track the temporal evolution of these themes over nearly two decades. This analysis reveals significant shifts in research priorities, such as the rising prominence of Alzheimer's research and healthcare outcomes, and highlights the correlation between network structure and thematic specialization.

The primary contribution of this work is the provision of a high-fidelity, feature-rich resource that bridges the gap between administrative metadata and structural network analysis. NIH-MPINet offers a standardized benchmark for developing and testing new statistical and machine learning algorithms, particularly those involving heterogeneous nodes, temporal dynamics, and text-embedded graphs. By making this dataset publicly available, we aim to empower researchers and policymakers to move beyond anecdotal observations to data-driven insights into the mechanisms that drive scientific discovery and team-based innovation.

\section{The NIH-MPINet Dataset}

Our analysis focuses on NIH R01-equivalent grants awarded to multiple PIs between 2006 and 2023, comprising 30,127 PIs and 86,743 grants. Notably, the multi-PI grant mechanism was formally introduced in 2006; thus, NIH-MPINet captures its entire historical trajectory. Grant-level data were retrieved from \href{https://reporter.nih.gov}{NIH RePORTER} and include attributes such as PI names, organizational affiliations, geographic locations, project titles, abstracts, NIH spending categories, keywords, and fiscal years. PI-level data were supplemented by scraping information from \href{https://pubmed.ncbi.nlm.nih.gov}{PubMed}. We detail the data collection and cleaning pipeline in Section \ref{sec:Data Collection}.

\begin{figure}[!htbp]
    \centering
    \includegraphics[width=\textwidth]{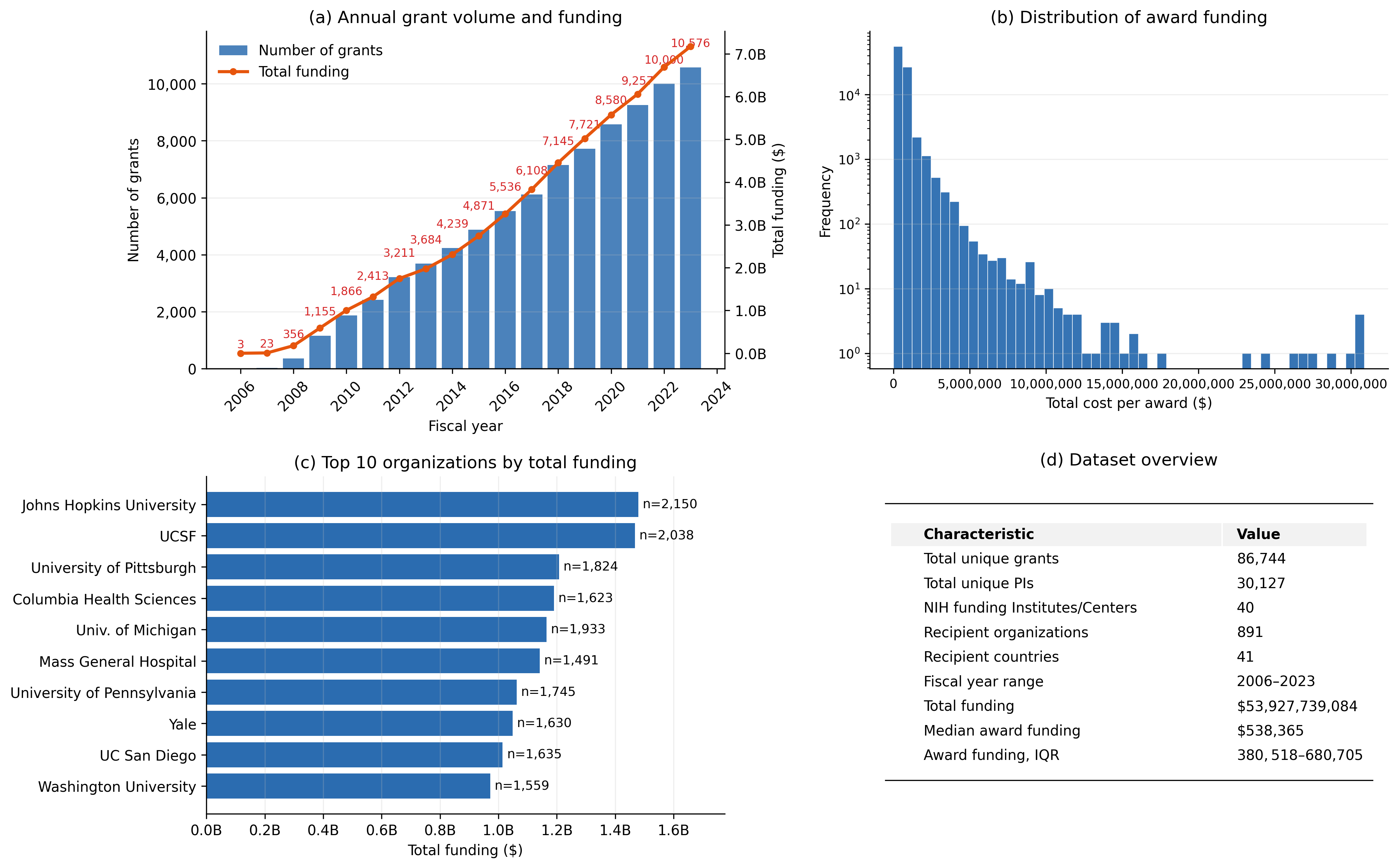}
    \caption{Descriptive summary of the multi-PI NIH R01-equivalent grant dataset.}
    \label{fig:summary_statistics}
\end{figure}

At the grant level, the dataset comprises 86,743 unique awards. We observe that both the annual volume of grants and total funding have increased steadily over time (Figure~\ref{fig:summary_statistics}a). These grants involve 30,127 PIs across 888 recipient organizations and are supported by 40 NIH Institutes and Centers (Figure~\ref{fig:summary_statistics}d). While the majority of awards are administered within the United States, recipient organizations span 41 countries. Funding distribution is highly concentrated, with a small subset of institutions accounting for a disproportionate share of total funding and grant counts (Figure~\ref{fig:summary_statistics}c). Award sizes exhibit a strongly right-skewed distribution, characterized by a high frequency of standard awards and a long tail of high-cost projects that contribute substantially to overall NIH expenditures (Figure~\ref{fig:summary_statistics}b).

At the PI level, we retrieved affiliation information for the 30,127 unique PIs. Successful data retrieval was achieved for 89\% of investigators; missing values are primarily due to webpage access constraints or limitations of the automated scraping algorithms. These PI-level attributes provide critical context regarding investigator backgrounds and institutional environments. Table \ref{tab:top10_pi_affiliation} summarizes the ten principal investigators with the highest number of NIH R01-equivalent multi-PI grants.

\begin{table}[!htbp]
\centering
\caption{Top 10 PIs by number of NIH R01-equivalent grants.}
\footnotesize
\renewcommand{\arraystretch}{1.0}

\resizebox{1.0\textwidth}{!}{  
\begin{tabular}{>{\RaggedRight\arraybackslash}p{4.5cm} c >{\RaggedRight\arraybackslash}p{7.5cm}}
\hline
\textbf{Principal Investigator} & \textbf{No. Grants} & \textbf{Affiliation} \\
\hline
CALHOUN, VINCE D. & 63 & Georgia State University; Georgia Institute of Technology; Emory University \\
WU, JOSEPH C. & 57 & Stanford University \\
SHEA, LONNIE D. & 54 & University of Michigan \\
CLARK, ANDREW G. & 47 & Cornell University \\
GIACOMINI, KATHLEEN M. & 43 & University of California, San Francisco \\
KAPLAN, DAVID L. & 43 & Tufts University \\
CASANOVA, JEAN-LAURENT & 41 & Rockefeller University; INSERM; University of Paris \\
BUCH, SHILPA J. & 41 & University of Nebraska Medical Center \\
GRINSTAFF, MARK W. & 39 & Boston University \\
WEISS, WILLIAM A. & 36 & University of California, San Francisco \\
\hline
\end{tabular}
}

\label{tab:top10_pi_affiliation}
\end{table}

\section{Data Curation and Network Construction}
\label{sec:Data Collection}

This section describes the methodological details underlying the curation of NIH-MPINet and the construction of the collaboration network.
\subsection{Grant (edge) -- level Data Extraction}

The grant-level information is derived from unique research projects identified by their administrative project numbers. Grant records were directly retrieved from the \href{https://reporter.nih.gov}{NIH RePORTER} database for the period 2006–2023. Using advanced search functionality, we filtered for multi-PI grants to specifically study collaborative research structures. Because R01-equivalent projects typically span multiple fiscal years, all longitudinal records associated with a project number were retained to ensure a complete representation of the collaborative effort.

\subsection{PI (node) -- level Feature Engineering}

The PI-level information was derived by identifying every unique investigator listed as a PI across the retrieved grant records. To obtain institutional metadata at the PI level, we leveraged auxiliary information beyond the multi-PI R01-equivalent dataset. For PIs also associated with single-PI grants, institutional metadata, including department and organizational affiliation, were extracted directly from NIH RePORTER. For PIs solely involved in multi-PI projects, we supplemented our database using affiliation information retrieved from \href{https://pubmed.ncbi.nlm.nih.gov}{PubMed}.

Specifically, we queried PubMed using the PI’s full name (first, middle, and last) to reduce potential name disambiguation errors and accessed the most recent publication (up to 2024). Affiliation strings were extracted by matching the investigator's name within the author list and parsing the corresponding institutional metadata. To ensure stability against API rate limits and high query volumes, the retrieval pipeline was implemented in Python using the \texttt{requests} and \texttt{BeautifulSoup} libraries with integrated batch-processing pauses.

\subsection{Network Representation}
We represent the curated data as a collaboration network $G = (V, E)$, where the set of nodes $V$ corresponds to the 30,127 unique PIs. An edge $(u, v) \in E$ is established between two PIs if they served as co-PIs on the same NIH R01-equivalent grant. Each edge represents at least one co-PI collaboration.


\section{Results}

\begin{figure}[!h]
	\centering
	\includegraphics[width=0.9\textwidth]{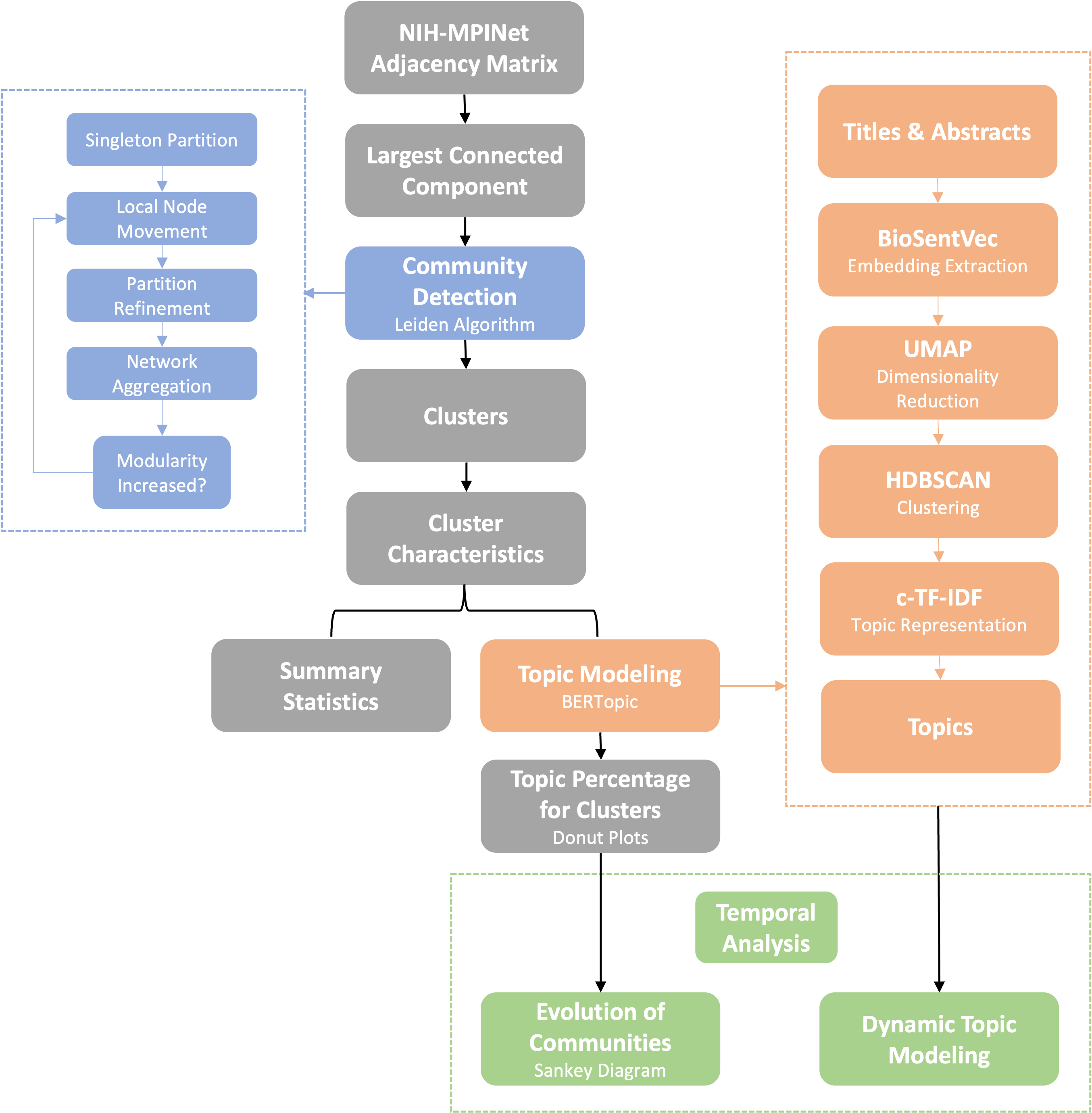}
	\caption{Flow diagram illustrating the overall analysis workflow. The left branch describes the construction of the PI collaboration network and community detection using the Leiden algorithm. The right branch shows the topic modeling process based on project titles and abstracts using BERTopic. The two branches are integrated to characterize community research focuses and investigate temporal trends.}
	\label{fig:flow_diagram}
\end{figure}

Our exploratory analysis aims to characterize collaboration structures among PIs, identify community-level patterns, and examine how research themes and collaborations evolve over time. To this end, we constructed a PI collaboration network from the multi-PI grant dataset, in which nodes represent PIs and edges indicate co-participation in the same grant. The network was represented as an unweighted adjacency matrix. An overview of the analytical workflow is shown in Figure~\ref{fig:flow_diagram}.

\begin{figure}[!htbp]
	\centering
	\begin{subfigure}[b]{0.46\textwidth}
		\centering
		\includegraphics[width=\textwidth]{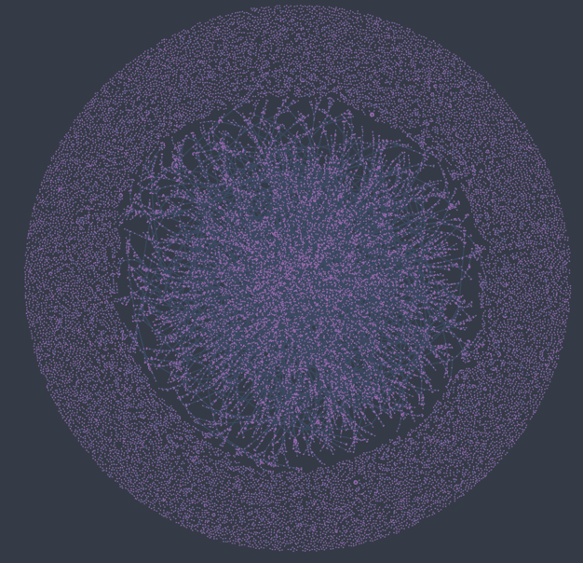}
		\caption{Full Network}
		\label{fig:lcc_full}
	\end{subfigure}
	\hfill
	\begin{subfigure}[b]{0.526\textwidth}
		\centering
		\includegraphics[width=\textwidth]{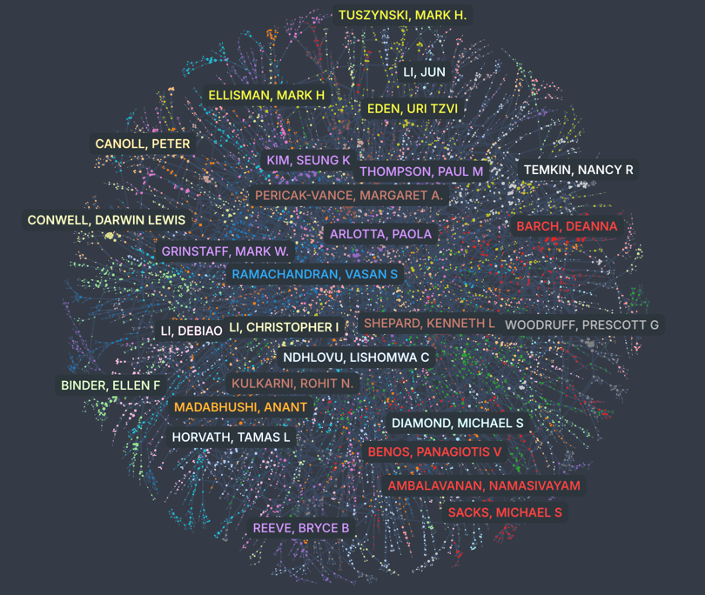}
		\caption{Largest Connected Component}
		\label{fig:cd_lcc}
	\end{subfigure}
	\caption{(a): The full network of NIH-MPINet; (b) the largest connected component, labeled by representative PIs colored by community labels.}
	\label{fig:lcc&cd}
\end{figure}

We focused subsequent analyses on the largest connected component of the PI collaboration network, defined as the largest subgraph in which any two PIs are connected through at least one path of collaborative relationships. It includes 13,873 PIs and 48,292 grants (Figure~\ref{fig:lcc_full}). The remaining components are substantially smaller and sparsely connected (Table~S1); therefore, restricting attention to the largest connected component is unlikely to affect the generality of the findings.

\subsection{Network Structure and Research Topics}

To understand the structural organization and thematic specialization of the multi-PI research landscape, we first identified the community structure within the largest connected component using the Leiden algorithm \citep{traag2019louvain}. This procedure revealed 19 distinct communities (Figure~\ref{fig:cd_lcc}), corresponding to groups of PIs with shared collaboration profiles. An interactive visualization of the network and its communities was generated using \href{\detokenize{https://cosmograph.app/run/?data=https://raw.githubusercontent.com/CuiranVivianShi/NIH_network/refs/heads/main/data/edges_largest_component.csv&meta=https://raw.githubusercontent.com/CuiranVivianShi/NIH_network/refs/heads/main/data/final_node_data.csv&source=source&target=target&gravity=0.25&repulsion=1&repulsionTheta=1.15&linkSpring=1&linkDistance=10&friction=0.85&renderLabels=true&renderHoveredLabel=true&renderLinks=true&nodeSizeScale=1&linkWidthScale=1&linkArrowsSizeScale=1&nodeSize=size-total%20links&nodeColor=color-node_color&linkWidth=width-default&linkColor=color-default&}}{Cosmograph}. Summary characteristics of these communities, including size and representative investigators, are provided in Table~\ref{tab:cluster_summary_main}, with additional details reported in the Supplementary Materials (Table S2).

\begin{table}[!htbp]
\scriptsize
\setlength{\tabcolsep}{2pt}
\renewcommand{\arraystretch}{1.0}
\centering
\caption{Summary of Identified 20 Topics}
\label{tab:topic_summary}

\begin{tabular}{c l c l}
\toprule
\textbf{Topic} & \textbf{Summary} & \textbf{Topic} & \textbf{Summary} \\
\midrule
T0  & Healthcare and Outcomes Research              & T10 & Virology: HIV and Viral Infections \\
T1  & Cellular Biology and Functions                & T11 & Pharmacology and Drug Development \\
T2  & Cardiovascular and Cognitive Health           & T12 & Alzheimer’s Disease and Neurodegenerative Disease \\
T3  & Neuroscience and Brain Disorders              & T13 & Infectious Disease in Developing Regions \\
T4  & Cancer Immunotherapy                          & T14 & Vaccines and Immune Response \\
T5  & Genomics and Genetic Variation                & T15 & Bone Health and Cancer \\
T6  & Molecular and Cellular Biology                & T16 & Ophthalmology and Vision Sciences \\
T7  & Neurology and Pain Management                 & T17 & Liver Diseases and Hepatic Function \\
T8  & Microbiome Research                           & T18 & Cellular Damage and DNA Repair \\
T9  & Medical Imaging Technologies                  & T19 & Sensory Biology and Ion Channels \\
\bottomrule
\end{tabular}

\end{table}

To characterize the research themes associated with each community, we applied topic modeling to grant titles and abstracts using BERTopic \citep{grootendorst2022bertopic} with BioSentVec embeddings \citep{chen2019biosentvec}. This approach yielded 20 topics capturing major thematic areas across the grant portfolio (Table \ref{tab:topic_summary}), based on the top ten keywords ranked by class-based c-TF-IDF importance and summarized into interpretable research themes using ChatGPT. While topic identification is data-driven through clustering and c-TF-IDF importance, the naming and interpretation of topics involve some degree of subjective judgment. Additional details, including topic-specific word frequency distributions and the topic similarity matrix, are shown in Figure S1 and S2, respectively. We also calculated topic distributions for each community and summarized the cluster characteristics accordingly (Table \ref{tab:cluster_summary_main}); the topic distributions for four representative communities are visualized using donut plots in Figure~\ref{fig:all_topic_dists}.

\begin{figure}[!htbp]
	\centering
	\includegraphics[width=1.0\textwidth]{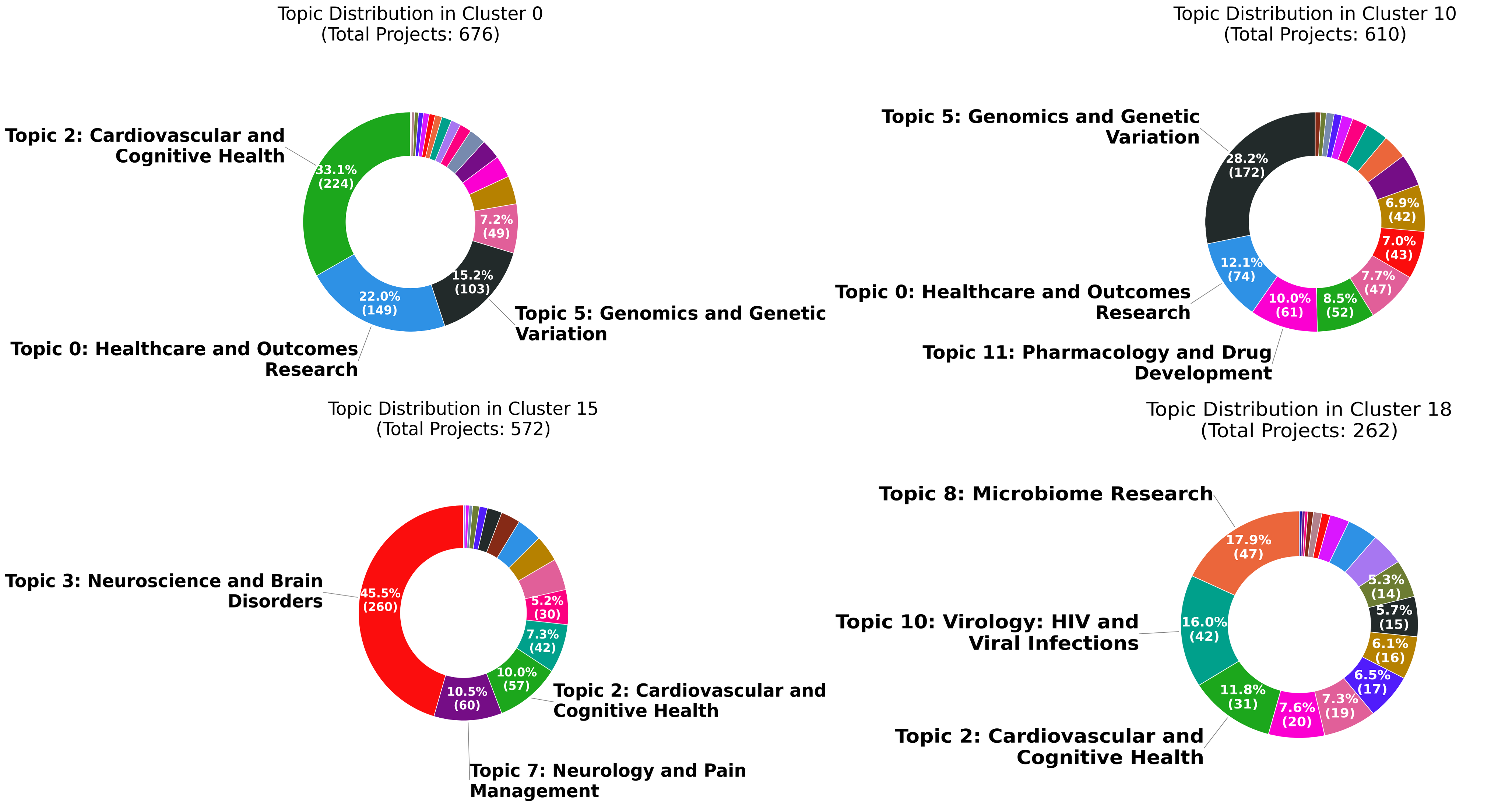}
\caption{Topic distribution across four selected NIH-MPINet clusters.}
\label{fig:all_topic_dists}
\end{figure}

\begin{table}[!htbp]
\centering
\caption{Summary of 19 Identified Clusters in NIH-MPINet}
\label{tab:cluster_summary_main}
\scriptsize
\renewcommand{\arraystretch}{1.00}

\begin{tabularx}{\textwidth}{
>{\raggedright\arraybackslash}p{0.8cm}
>{\centering\arraybackslash}p{1.5cm}
>{\raggedright\arraybackslash}p{3.2cm}
>{\raggedright\arraybackslash}p{3.6cm}
>{\raggedright\arraybackslash}X
}
\toprule
\textbf{Cluster} & \textbf{Nodes / Edges} & \textbf{Top PI} & \textbf{Top Topics (\%)} & \textbf{Summary} \\
\midrule

C0  & 847 / 1295 & RAMACHANDRAN, VASAN S & T2 (33.1\%), T0 (22.0\%), T5 (15.2\%) & Cardiovascular Health, Outcomes Research, Genomics \\
C1  & 841 / 1131 & SCAMPAVIA, LOUIS DANIEL & T0 (30.9\%), T11 (12.9\%), T10 (11.4\%) & Outcomes Research, Drug Development, Virology \\
C2  & 779 / 1123 & MADABHUSHI, ANANT & T2 (20.7\%), T0 (16.9\%), T4 (8.0\%) & Cardiovascular Health, Outcomes Research, Cancer \\
C3  & 778 / 1041 & GRAY, NATHANAEL SCHIANDER & T4 (20.2\%), T1 (13.2\%), T6 (11.8\%) & Cancer, Cell Biology, Molecular Biology \\
C4  & 776 / 1141 & XU, XIANGMIN & T3 (20.6\%), T1 (11.1\%), T0 (8.5\%) & Neuroscience, Cell Biology, Outcomes Research \\
C5  & 769 / 1115 & ZHANG, RONG & T5 (13.3\%), T4 (12.2\%), T1 (11.1\%) & Genomics, Cancer, Cell Biology \\
C6  & 766 / 1228 & BOOKHEIMER, SUSAN Y & T3 (20.5\%), T0 (18.2\%), T9 (12.7\%) & Brain Disorders, Outcomes Research, Imaging \\
C7  & 762 / 1089 & CHUNG, WENDY K & T2 (26.8\%), T0 (24.3\%), T5 (10.3\%) & Cardiovascular Health, Outcomes Research, Genomics \\
C8  & 743 / 1210 & POWERS, ALVIN C & T0 (29.8\%), T1 (16.3\%), T5 (12.3\%) & Outcomes Research, Cell Biology, Genomics \\
C9  & 739 / 1196 & PERICAK-VANCE, MARGARET A. & T5 (18.2\%), T1 (16.9\%), T3 (15.7\%) & Genomics, Cell Biology, Neuroscience \\
C10 & 733 / 1148 & KENNY, EIMEAR ELIZABETH & T5 (28.2\%), T0 (12.1\%), T11 (10.0\%) & Genomics, Outcomes Research, Pharmacology \\
C11 & 723 / 1046 & DEBAUN, MICHAEL R. & T1 (31.5\%), T4 (12.6\%), T6 (11.9\%) & Cell Biology, Cancer, Molecular Biology \\
C12 & 723 / 1091 & LI, DEBIAO & T1 (20.0\%), T2 (15.2\%), T5 (14.6\%) & Cell Biology, Cardiovascular Health, Genomics \\
C13 & 716 / 1097 & WOODRUFF, PRESCOTT G & T13 (14.8\%), T1 (12.9\%), T4 (11.7\%) & Infectious Disease, Cell Biology, Cancer \\
C14 & 715 / 1207 & PLEVRITIS, SYLVIA KATINA & T0 (40.0\%), T2 (18.7\%), T3 (13.7\%) & Outcomes Research, Cardiovascular Health, Neuroscience \\
C15 & 713 / 1169 & ELLISMAN, MARK H & T3 (45.5\%), T7 (10.5\%), T2 (10.0\%) & Neuroscience, Pain, Cardiovascular Health \\
C16 & 711 / 1092 & FIGUEIREDO, JANE C. & T5 (31.9\%), T0 (17.2\%), T2 (10.0\%) & Genomics, Outcomes Research, Cardiovascular Health \\
C17 & 705 / 989  & LEWIS, JASON S. & T0 (22.7\%), T3 (18.3\%), T4 (10.3\%) & Outcomes Research, Neuroscience, Cancer \\
C18 & 334 / 448  & DIAMOND, MICHAEL S & T8 (17.9\%), T10 (16.0\%), T2 (11.8\%) & Microbiome, Virology, Cardiovascular Health \\

\bottomrule
\end{tabularx}
\end{table}

\subsection{Temporal Trends in Communities and Topics}
Since research priorities and team configurations are inherently dynamic, we investigated the temporal evolution of the network to characterize longitudinal shifts in themes and collaborative structures. We first used the dynamic BERTopic model to study how topics changed over time. This model updates topic representations at each time step, allowing us to track topic evolution without needing to rerun the full model for each
interval. Topic trajectories reveal differential growth and decline among major research areas (Figures~\ref{fig:time_visualization_sankey_diagram}a and S3-S4). In 2006, only Topic~1 (\textit{Cellular Biology and Functions}) and Topic~4 (\textit{Cancer Immunotherapy}) were present, with other topics emerging gradually. 


 Several significant trends emerge from the temporal analysis. We observe a distinct rise in Topic 0 (Healthcare and Outcomes Research) beginning in 2018, as well as steady increases in Topic 2 (Cardiovascular and Cognitive Health) and Topic 12 (Alzheimer's Disease and Neurodegenerative Disease) since 2016. Conversely, foundational biological topics have seen a relative decline in multi-PI representation, with Topic 1 (Cellular Biology and Functions) and Topic 6 (Molecular and Cellular Biology) generally trending downward since 2010.



\begin{figure}[!htbp]
    \centering
    \includegraphics[width=0.8\textwidth]{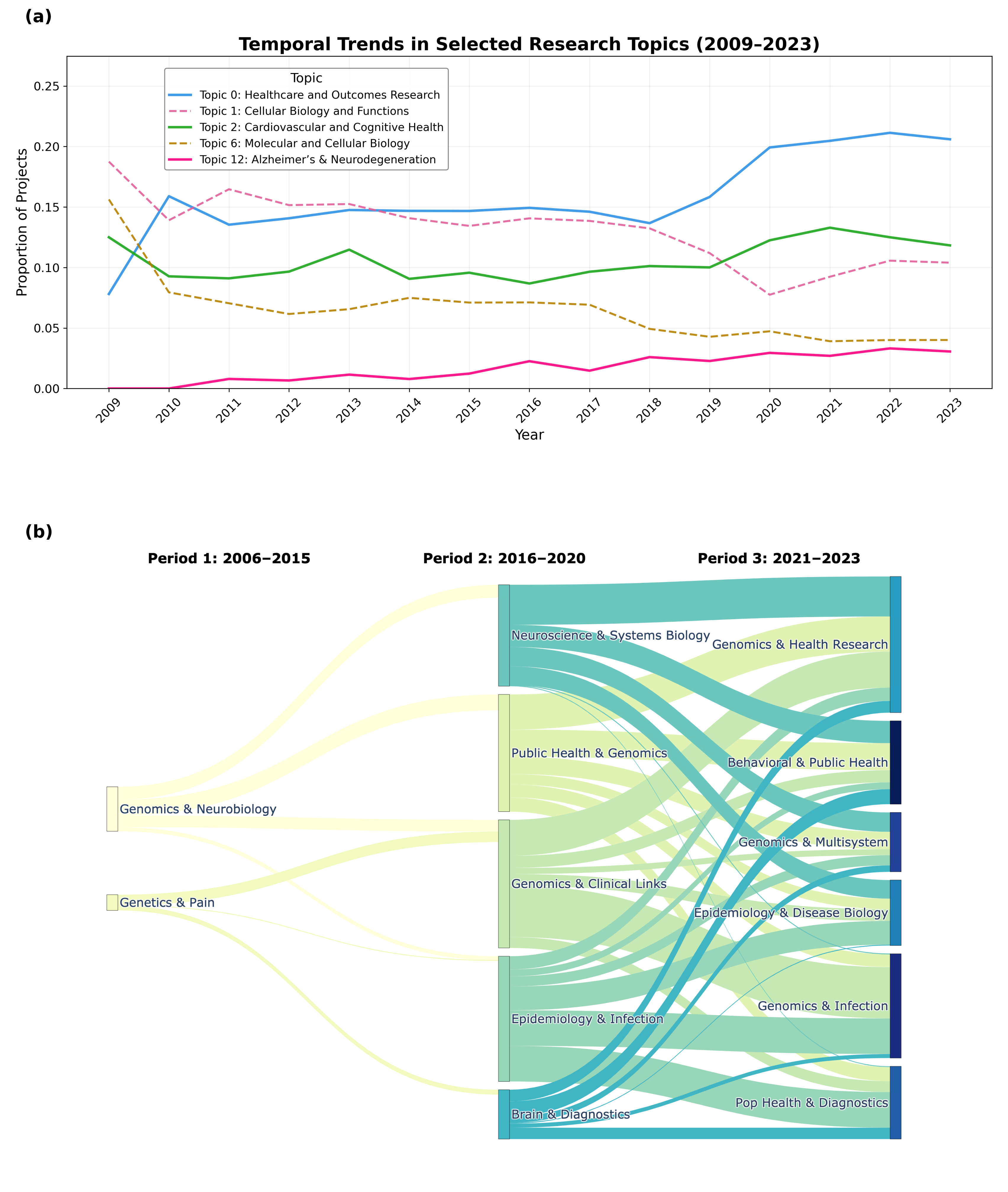}
    
    \caption{Temporal dynamics and structural evolution of research topics.
    (a) Temporal trends in five selected research topics (2009--2023), showing the relative proportion of projects across key domains. Trends are shown from 2008 onward because the number of multi-PI projects in 2006–2007 was too small for stable topic identification (see Figures S3–S4).
    (b) Sankey diagram illustrating the evolution of collaboration clusters across time periods. Flow widths are proportional to the number of shared PIs between clusters in adjacent periods.
    }
    \label{fig:time_visualization_sankey_diagram}
\end{figure}

Additionally, we stratified the datasets into three periods: 2006--2015, 2016--2020, and 2021--2023. We initially considered 5-year intervals, but the number of multi-PI projects prior to 2015 was relatively small (Figure \ref{fig:summary_statistics}a), so earlier years were combined to ensure sufficient sample size within each period. We applied the Leiden algorithm separately to each subset, showing evolving community structures. The number of identified clusters increased over time: two in 2006–2015, five in 2016–2020, and six in 2021–2023. Topic modeling was then performed on each set of clusters to characterize their thematic focus. To visualize these dynamics, we constructed a Sankey diagram (Figure~\ref{fig:time_visualization_sankey_diagram}b) to illustrate the transformation of communities based on overlapping PIs, where flow widths are proportional to the number of shared PIs between clusters in adjacent periods. The diagram highlights the sustained prominence of genetics and genomics. Notably, many projects originally associated with \textit{Neuroscience and Systems Biology} (2006–2020) transitioned into clusters centered on \textit{Genomics and Health Research} in 2021–2023. Recently, a prominent \textit{Genomics and Infection} cluster emerged, aggregating projects previously distributed across \textit{Genomics and Clinical Links} and \textit{Epidemiology and Infection} (2016–2020).

These transitions likely reflect major shifts in the national research agenda and institutional funding priorities. For instance, the migration of investigators from \textit{Neuroscience and Systems Biology} toward \textit{Health Research} suggests a shift from fundamental biological mapping toward clinical and translational applications, particularly in population health and precision medicine. Similarly, the emergence of the \textit{Genomics and Infection} cluster in the 2021–2023 period directly corresponds to the NIH's intensive pivot toward COVID-19 research and pandemic preparedness, which necessitated a massive integration of high-throughput genomic surveillance with infectious disease modeling.  These patterns demonstrate how the multi-PI mechanism allows the research community to rapidly reorganize its structural topology in response to emerging global health crises and evolving technological frontiers.

\section{Discussion and Future Work}

The primary contribution of this article is the curation and release of the NIH-MPINet dataset, a high-fidelity resource representing nearly two decades of multi-PI collaboration on NIH R01-equivalent grants. This dataset is uniquely valuable because it bridges the gap between raw administrative metadata and structural network analysis, providing a unified framework that integrates grant-level textual information with investigator-level institutional attributes. By removing the significant technical barriers associated with manual data extraction and entity resolution from disconnected platforms, NIH-MPINet empowers the research community to conduct large-scale, reproducible studies on the mechanisms that drive scientific discovery and team-based innovation.

Our exploratory analysis demonstrated the utility of NIH-MPINet for uncovering both the structural organization and thematic evolution of biomedical research. Using community detection and topic modeling, we identified specialized collaborative hubs, ranging from neuroscience to cancer immunotherapy, and tracked their reorganization over time. Notably, our analysis revealed significant longitudinal shifts, such as the rising prominence of Alzheimer's research and healthcare outcomes, and the structural pivot of the research community toward pandemic preparedness following the COVID-19 crisis. These findings illustrate how multi-PI mechanisms allow the scientific enterprise to rapidly reorganize its topology in response to emerging global health priorities.

Moving forward, we expect NIH-MPINet to serve as a foundational benchmark for advancing the field of Team Science and the development of specialized statistical learning methods for heterogeneous, text-embedded graphs. The availability of such high-resolution funding data supports a transition from anecdotal observations of research trends to rigorous, data-driven modeling of scientific collaboration. By making this resource public, we aim to push forward research into funding dynamics, interdisciplinary team assembly, and the long-term impact of research policies on scientific productivity and innovation. Future work based on NIH-MPINet could proceed in several key directions:
\begin{itemize}

\item Economic Integration: Incorporating grant-level funding amounts and fiscal variables to model the relationship between financial investment and collaborative network topology.

\item Career Dynamics: Integrating investigator career stages and funding histories to analyze team composition and the influence of early-career vs. senior PIs on research directions.

\item Impact Mapping: Linking the grant-based collaboration network to downstream research outputs, such as citation counts and patent filings, to clarify the relationship between structural collaboration and scientific impact.

\item Multi-layer Expansion: Constructing multi-layer networks that simultaneously represent grant collaborations, co-authorship records, and co-citation patterns for a more holistic view of team science.

\item Algorithmic Benchmarking: Utilizing the dataset as a standardized resource for testing and refining dynamic graph neural networks and other emerging statistical learning methods on real-world, high-dimensional scientific data.
\end{itemize}

\bibliographystyle{plainnat} 
\bibliography{references} 

@article{talley2011database,
	title={Database of NIH grants using machine-learned categories and graphical clustering},
	author={Talley, Edmund M and Newman, David and Mimno, David and Herr, Bruce W and Wallach, Hanna M and Burns, Gully APC and Leenders, AG Miriam and McCallum, Andrew},
	journal={Nature Methods},
	volume={8},
	number={6},
	pages={443--444},
	year={2011},
	publisher={Nature Publishing Group US New York}
}

@article{zhang2023turtling,
	title={Turtling: a time-aware neural topic model on NIH grant data},
	author={Zhang, Ruiyi and Duan, Ziheng and Lee, CheYu and Riffle, Dylan and Min, Martin Renqiang and Zhang, Jing},
	journal={Bioinformatics Advances},
	volume={3},
	number={1},
	pages={vbad096},
	year={2023},
	publisher={Oxford University Press}
}

@article{ji2022co,
	title={Co-citation and co-authorship networks of statisticians},
	author={Ji, Pengsheng and Jin, Jiashun and Ke, Zheng Tracy and Li, Wanshan},
	journal={Journal of Business \& Economic Statistics},
	volume={40},
	number={2},
	pages={469--485},
	year={2022},
	publisher={Taylor \& Francis}
}

@article{zhu2023novel,
  title={A novel NIH research grant recommender using BERT},
  author={Zhu, Jie and Patra, Braja Gopal and Wu, Hulin and Yaseen, Ashraf},
  journal={PloS one},
  volume={18},
  number={1},
  pages={e0278636},
  year={2023},
  publisher={Public Library of Science San Francisco, CA USA}
}

@article{franceschi2017patterns,
  title={Patterns of recent National Institutes of Health (NIH) funding to diagnostic radiology departments: analysis using the NIH RePORTER system},
  author={Franceschi, Ana M and Rosenkrantz, Andrew B},
  journal={Academic radiology},
  volume={24},
  number={9},
  pages={1162--1168},
  year={2017},
  publisher={Elsevier}
}

@inproceedings{malikireddy2017network,
  title={Network analysis of NIH grant critiques},
  author={Malikireddy, Dastagiri Reddy and Jens, Madeline and Filut, Amarette and Bhattacharya, Anupama and Pier, Elizabeth L and Lee, You Geon and Carnes, Molly and Kaatz, Anna},
  booktitle={Proceedings of the 2017 IEEE/ACM International Conference on Advances in Social Networks Analysis and Mining 2017},
  pages={240--243},
  year={2017}
}

@article{traag2019louvain,
	title={From Louvain to Leiden: guaranteeing well-connected communities},
	author={Traag, Vincent A and Waltman, Ludo and Van Eck, Nees Jan},
	journal={Scientific reports},
	volume={9},
	number={1},
	pages={1--12},
	year={2019},
	publisher={Nature Publishing Group}
}

@article{grootendorst2022bertopic,
	title={BERTopic: Neural topic modeling with a class-based TF-IDF procedure},
	author={Grootendorst, Maarten},
	journal={arXiv preprint arXiv:2203.05794},
	year={2022}
}

@inproceedings{chen2019biosentvec,
	title={BioSentVec: creating sentence embeddings for biomedical texts},
	author={Chen, Qingyu and Peng, Yifan and Lu, Zhiyong},
	booktitle={2019 IEEE International Conference on Healthcare Informatics (ICHI)},
	pages={1--5},
	year={2019},
	organization={IEEE}
}

@article{national2015enhancing,
  title={Enhancing the effectiveness of team science},
  author={National Research Council and others},
  journal={National Research Council. 2015. Enhancing the Effectiveness of Team Science. Washington},
  pages={19007},
  year={2015}
}

@misc{nih_budget,
  author = {{National Institutes of Health}},
  title = {Budget},
  howpublished = {\url{https://www.nih.gov/about-nih/organization/budget}},
  year = {2024},
  note = {Accessed: 2026-03-22}
}



\section*{Appendix}

\subsection*{Appendix A: Data and Code Availability}The NIH-MPINet dataset is available at 
\url{https://huggingface.co/datasets/LiLabUNC/NIH-MPINet}.
All analysis code is available at 
\url{https://github.com/CuiranVivianShi/NIH_network}.

\subsection*{Appendix B: Additional tables and figures}
\setcounter{table}{0}
\renewcommand{\thetable}{S\arabic{table}}

\setcounter{figure}{0}
\renewcommand{\thefigure}{S\arabic{figure}}

\textbf{Connected component size distribution.}
Table~S1 summarizes the size distribution of connected components in the NIH-MPINet collaboration network. Community detection was performed using the Leiden algorithm with the RBConfigurationVertexPartition objective function, resolution parameter set to 0.0001, maximum community size of 850, 20 iterations, and a fixed random seed of 42.
The network is highly fragmented, with most components consisting of only a small number of investigators (median size = 2). 
Specifically, 3,383 components include two investigators, and 1,115 include three investigators. 
In contrast, one large connected component includes 13,873 investigators and represents the main collaboration structure of the network. 
Overall, the network includes one large collaboration group along with many smaller, separate teams.

\begin{table}[H]
\centering
\caption{
Connected component size characteristics in the NIH-MPINet collaboration network.
Most components are small (median size = 2), while one giant component contains 13,873 investigators.
}
\label{tab:component_size}

\small
\renewcommand{\arraystretch}{1.15}

\begin{tabular}{lc}
\hline
\textbf{Summary statistic} & \textbf{Value} \\
\hline
Maximum component size & 13,873 \\
Minimum component size & 2 \\
Mean component size & 5.50 \\
Median component size & 2 \\
\hline
\multicolumn{2}{l}{\textbf{Component size distribution}} \\
\hline
Component size & Number of components \\
\hline
13,873 & 1 \\
53 & 1 \\
35 & 1 \\
34 & 1 \\
33 & 1 \\
30 & 1 \\
27 & 1 \\
24 & 1 \\
23 & 2 \\
22 & 1 \\
21 & 1 \\
20 & 4 \\
19 & 1 \\
18 & 4 \\
17 & 6 \\
16 & 4 \\
15 & 8 \\
14 & 7 \\
13 & 12 \\
12 & 12 \\
11 & 17 \\
10 & 33 \\
9 & 27 \\
8 & 48 \\
7 & 75 \\
6 & 117 \\
5 & 212 \\
4 & 383 \\
3 & 1,115 \\
2 & 3,383 \\
\hline
\end{tabular}

\end{table}

\clearpage
\textbf{Community detection results.}
Table~S2 summarizes the characteristics of the communities identified in the NIH-MPINet collaboration network using the Leiden algorithm. 
For each cluster, we report the number of investigators (nodes), the number of collaborations (edges), and the average node degree. 
Most clusters are similar in size, generally including about 700–850 investigators, which shows a relatively balanced partition of the main connected component. 
The table also lists the top 10 representative principal investigators in each cluster based on normalized degree.

\begin{scriptsize}
\begin{longtable}{>{\raggedright\arraybackslash}p{1cm} >{\raggedright\arraybackslash}p{2.6cm} >{\raggedright\arraybackslash}p{2.2cm} >{\raggedright\arraybackslash}p{8.2cm}}
\caption{
Summary of community detection results in the NIH-MPINet collaboration network. 
For each cluster, the table reports the number of nodes and edges, average node degree, and the top 10 representative principal investigators ranked by normalized degree.
}
\label{tab:cluster_summary_combined} \\
\toprule
\textbf{Cluster} & \textbf{Nodes / Edges} & \textbf{Average Node Degree} & \textbf{Top 10 Representative PIs (Normalized Degrees)} \\
\midrule
\endfirsthead

\toprule
\textbf{Cluster} & \textbf{Nodes / Edges} & \textbf{Average Node Degree} & \textbf{Top 10 Representative PIs (Normalized Degrees)} \\
\midrule
\endhead

\midrule
\multicolumn{4}{r}{\textit{Continued on next page}} \\
\endfoot

\bottomrule
\endlastfoot

1 & 847 / 1295 & 3.06 & \makecell[l]{
RAMACHANDRAN, VASAN S (6.30) \\
SESHADRI, SUDHA (6.30) \\
MESCHIA, JAMES F (5.04) \\
GERSZTEN, ROBERT E (5.04) \\
HOWARD, GEORGE (4.61) \\
BRODERICK, JOSEPH PAUL (4.19) \\
PSATY, BRUCE M (4.19) \\
PERREIRA, KRISTA M (3.77) \\
KOTTON, DARRELL N. (3.35) \\
SACCO, RALPH L. (3.35)} \\
\midrule

2 & 841 / 1131 & 2.69 & \makecell[l]{
SCAMPAVIA, LOUIS DANIEL (8.80) \\
NDHLOVU, LISHOMWA C (7.27) \\
HORVATH, TAMAS L (5.24) \\
SPRINGER, SANDRA ANN (4.73) \\
ROSENBERG, PAUL B (4.73) \\
CARRICO, ADAM WAYNE (4.73) \\
BUCH, SHILPA J. (3.71) \\
IWASAKI, AKIKO (3.71) \\
MIMIAGA, MATTHEW JAMES (3.21) \\
SALTZMAN, W. MARK (3.21)} \\
\midrule

3 & 779 / 1123 & 2.88 & \makecell[l]{
MADABHUSHI, ANANT (7.68) \\
OGEDEGBE, OLUGBENGA G. (5.30) \\
MASON, CHRISTOPHER EDWARD (4.82) \\
NIZET, VICTOR (4.82) \\
WANG, GE (4.82) \\
HOU, LIFANG (4.35) \\
EZECHI, OLIVER CHUKWUJEKWU (3.87) \\
NAVAS-ACIEN, ANA (3.87) \\
LARIN, KIRILL V (3.87) \\
ALCAIDE, MARIA LUISA (3.87)} \\
\midrule

4 & 778 / 1041 & 2.68 & \makecell[l]{
GRAY, NATHANAEL SCHIANDER (6.32) \\
WILSON, DAVID M (5.30) \\
LIN, HENING (4.78) \\
WU, HAO (4.27) \\
FELDMAN, EVA LUCILLE (4.27) \\
WANG, YI (4.27) \\
SUH, YOUSIN (4.27) \\
CHENG, FEIXIONG (3.76) \\
YU, HAIYUAN (3.76) \\
KARCZMAR, GREGORY S. (3.24)} \\
\midrule

5 & 776 / 1141 & 2.94 & \makecell[l]{
XU, XIANGMIN (6.88) \\
YEO, EUGENE WEI-MING (5.97) \\
REN, BING (5.51) \\
MILLER, LEE (5.51) \\
GOLSHANI, PEYMAN (3.68) \\
CALHOUN, VINCE D (3.68) \\
ZHANG, KUN (3.68) \\
CHEN, HONG (3.68) \\
WANG, LEI (3.23) \\
XING, YI (3.23)} \\
\midrule

6 & 769 / 1115 & 2.90 & \makecell[l]{
ZHANG, RONG (5.86) \\
BINDER, ELLEN F (5.86) \\
IRIMIA, DANIEL (4.41) \\
STEINBACH, WILLIAM J (4.41) \\
BATRA, SURINDER K. (4.41) \\
MARIUZZA, ROY A (3.92) \\
BURNS, JEFFREY MURRAY (3.44) \\
SEGRE, DANIEL (3.44) \\
DEWHIRST, FLOYD E (3.44) \\
LI, WEI (3.44)} \\
\midrule

7 & 766 / 1228 & 3.21 & \makecell[l]{
BOOKHEIMER, SUSAN Y (5.93) \\
BARCH, DEANNA (5.93) \\
VAN ESSEN, DAVID C (5.12) \\
BENOS, PANAGIOTIS V (4.72) \\
BUCKNER, RANDY L (4.32) \\
AMBALAVANAN, NAMASIVAYAM (4.32) \\
SMITH, STEPHEN MARK (3.92) \\
HEWITT, JOHN K. (3.92) \\
DAPRETTO, MIRELLA (3.92) \\
UGURBIL, KAMIL (3.52)} \\
\midrule

8 & 762 / 1089 & 2.86 & \makecell[l]{
CHUNG, WENDY K (7.21) \\
KIZER, JORGE R (6.19) \\
HOROWITZ, CAROL R (4.15) \\
NEEDHAM, DALE MURRAY (3.64) \\
HOUGH, CATHERINE LEE (3.64) \\
KRITCHEVSKY, STEPHEN B. (3.64) \\
NEWMAN, ANNE B. (3.13) \\
WAPNER, RONALD (3.13) \\
IX, JOACHIM H (3.13) \\
WENG, CHUNHUA (3.13)} \\
\midrule

9 & 743 / 1210 & 3.26 & \makecell[l]{
POWERS, ALVIN C (8.78) \\
KIM, SEUNG K (5.02) \\
KOENEN, KARESTAN C (5.02) \\
RESSLER, KERRY J. (4.34) \\
THOMPSON, PAUL M (4.34) \\
NEALE, BENJAMIN MICHAEL (4.00) \\
SUNYAEV, SHAMIL (4.00) \\
MACDONALD, PATRICK (3.66) \\
BRISSOVA, MARCELA (3.66) \\
ATKINSON, MARK A. (3.66)} \\
\midrule

10 & 739 / 1196 & 3.24 & \makecell[l]{
PERICAK-VANCE, MARGARET A. (7.75) \\
HAROUTUNIAN, VAHRAM (5.61) \\
ROUSSOS, PANAGIOTIS (5.61) \\
BLANGERO, JOHN (5.61) \\
SHEPARD, KENNETH L (4.90) \\
ZHANG, BIN (4.54) \\
HAINES, JONATHAN L (4.19) \\
SCHADT, ERIC E (4.19) \\
LOIS, CARLOS (4.19) \\
SESTAN, NENAD (3.47)} \\
\midrule

11 & 733 / 1148 & 3.14 & \makecell[l]{
KENNY, EIMEAR ELIZABETH (6.52) \\
EICHLER, EVAN (5.28) \\
CLARK, ANDREW G (4.87) \\
MCCARTHY, MARK IAN (4.46) \\
WANG, TING (4.46) \\
GESCHWIND, DANIEL H (4.05) \\
MONTINE, THOMAS J (3.64) \\
LIPTON, RICHARD B. (3.23) \\
FREIMER, NELSON B. (3.23) \\
DABELEA, DANA (3.23)} \\
\midrule

12 & 723 / 1046 & 2.89 & \makecell[l]{
DEBAUN, MICHAEL R. (7.19) \\
WEISS, MITCHELL J (4.81) \\
SNYDER, MICHAEL P. (4.33) \\
STEIN, LINCOLN D. (3.86) \\
STAMER, W DANIEL (3.86) \\
TRAYANOVA, NATALIA A. (3.86) \\
SEEWALDT, VICTORIA L. (3.86) \\
WU, JOSEPH C. (3.86) \\
KHADEMHOSSEINI, ALI (3.38) \\
BLOBEL, GERD A (3.38)} \\
\midrule

13 & 723 / 1091 & 3.02 & \makecell[l]{
LI, DEBIAO (7.05) \\
ZHANG, JIANYI (4.70) \\
LENBURG, MARC ELLIOTT (4.70) \\
LEIN, PAMELA J (4.70) \\
SPIRA, AVRUM E (4.70) \\
ABERLE, DENISE R. (4.23) \\
ORALKAN, OMER (3.76) \\
PAGE, JR., C DAVID (2.82) \\
SHIM, HYUNSUK (2.82) \\
WONG, DAVID T (2.82)} \\
\midrule

14 & 716 / 1097 & 3.07 & \makecell[l]{
WOODRUFF, PRESCOTT G (5.68) \\
WENZEL, SALLY E (5.68) \\
LEVY, BRUCE D (4.86) \\
HOFFMAN, ERIC ALFRED (4.46) \\
CASTRO, MARIO (4.46) \\
ISRAEL, ELLIOT (4.46) \\
MEYERS, DEBORAH A. (4.46) \\
DENLINGER, LOREN C (4.05) \\
MAUGER, DAVID T. (4.05) \\
COMHAIR, SUZY (4.05)} \\
\midrule

15 & 715 / 1207 & 3.38 & \makecell[l]{
PLEVRITIS, SYLVIA KATINA (6.06) \\
DE KONING, HARRY J (6.06) \\
MANDELBLATT, JEANNE (5.34) \\
TEMKIN, NANCY R (4.97) \\
KONG, CHUNG YIN (4.61) \\
HAAS, JENNIFER S (3.88) \\
JIANG, XIAOQIAN (3.51) \\
BIAN, JIANG (3.15) \\
REIMAN, ERIC MICHAEL (3.15) \\
DIAZ-ARRASTIA, RAMON (3.15)} \\
\midrule

16 & 713 / 1169 & 3.28 & \makecell[l]{
ELLISMAN, MARK H (5.62) \\
DICKERSON, BRADFORD C. (5.24) \\
LIANGPUNSAKUL, SUTHAT (5.24) \\
SEUNG, HYUNJUNE SEBASTIAN (4.86) \\
FOROUD, TATIANA M. (4.86) \\
EDEN, URI TZVI (4.48) \\
JOHNSON, STERLING C (3.71) \\
KARN, JONATHAN (3.33) \\
CHEN, WEI (3.33) \\
KRAMER, MARK ALAN (3.33)} \\
\midrule

17 & 711 / 1092 & 3.07 & \makecell[l]{
FIGUEIREDO, JANE C. (6.76) \\
LE MARCHAND, LOIC (5.16) \\
HAIMAN, CHRISTOPHER ALAN (4.77) \\
NORTH, KARI E. (4.37) \\
CONWELL, DARWIN LEWIS (3.97) \\
ULRICH, CORNELIA M (3.97) \\
BELLIN, MELENA D. (3.97) \\
LI, CHRISTOPHER I (3.57) \\
STEEN, HANNO (3.57) \\
HART, PHILIP A. (3.57)} \\
\midrule

18 & 705 / 989 & 2.81 & \makecell[l]{
LEWIS, JASON S. (6.40) \\
SHAW, DANIEL S (5.35) \\
TERRY, MARY BETH (4.83) \\
GENNARO, MARIA LAURA (4.30) \\
FAYAD, ZAHI A. (3.78) \\
GUO, PEIXUAN (3.78) \\
CANINO, GLORISA J (3.25) \\
MONK, CATHERINE E (3.25) \\
BRITTON, ROBERT A (3.25) \\
ROCKNE, RUSSELL CHRISTIAN (3.25)} \\
\midrule

19 & 334 / 448 & 2.68 & \makecell[l]{
DIAMOND, MICHAEL S (8.77) \\
LI, JUN (4.21) \\
FREMONT, DAVED H. (3.20) \\
VAN VOORHIS, WESLEY C (3.20) \\
STAPPENBECK, THADDEUS S (3.20) \\
GARCIA-SASTRE, ADOLFO (2.69) \\
DANTAS, GAUTAM (2.69) \\
XAVIER, RAMNIK J (2.69) \\
PADMANABHAN, VASANTHA (2.69) \\
WANG, YONG (2.19)} \\
\end{longtable}
\end{scriptsize}

\clearpage
\textbf{Topic keyword representation.}
Figure~S1 shows the top keywords for each topic identified using BERTopic based on BioSentVec embeddings, with keywords ranked by c-TF-IDF importance. 
These keywords reflect major biomedical research areas represented in the NIH-MPINet collaboration network. 
Based on these keyword patterns, topics were summarized into interpretable research themes in Table~2, including healthcare outcomes (Topic~0), cardiovascular and cognitive health (Topic~2), neuroscience (Topic~3), genomics (Topic~5), microbiome research (Topic~8), virology (Topic~10), pharmacology (Topic~11), and medical imaging (Topic~9).
Across the full corpus, 32\% of grants were assigned to the outlier category (Topic -1), which indicates they did not strongly align with any single topic.

\begin{figure}[htbp]
	\centering
	\includegraphics[width=0.8\textwidth]{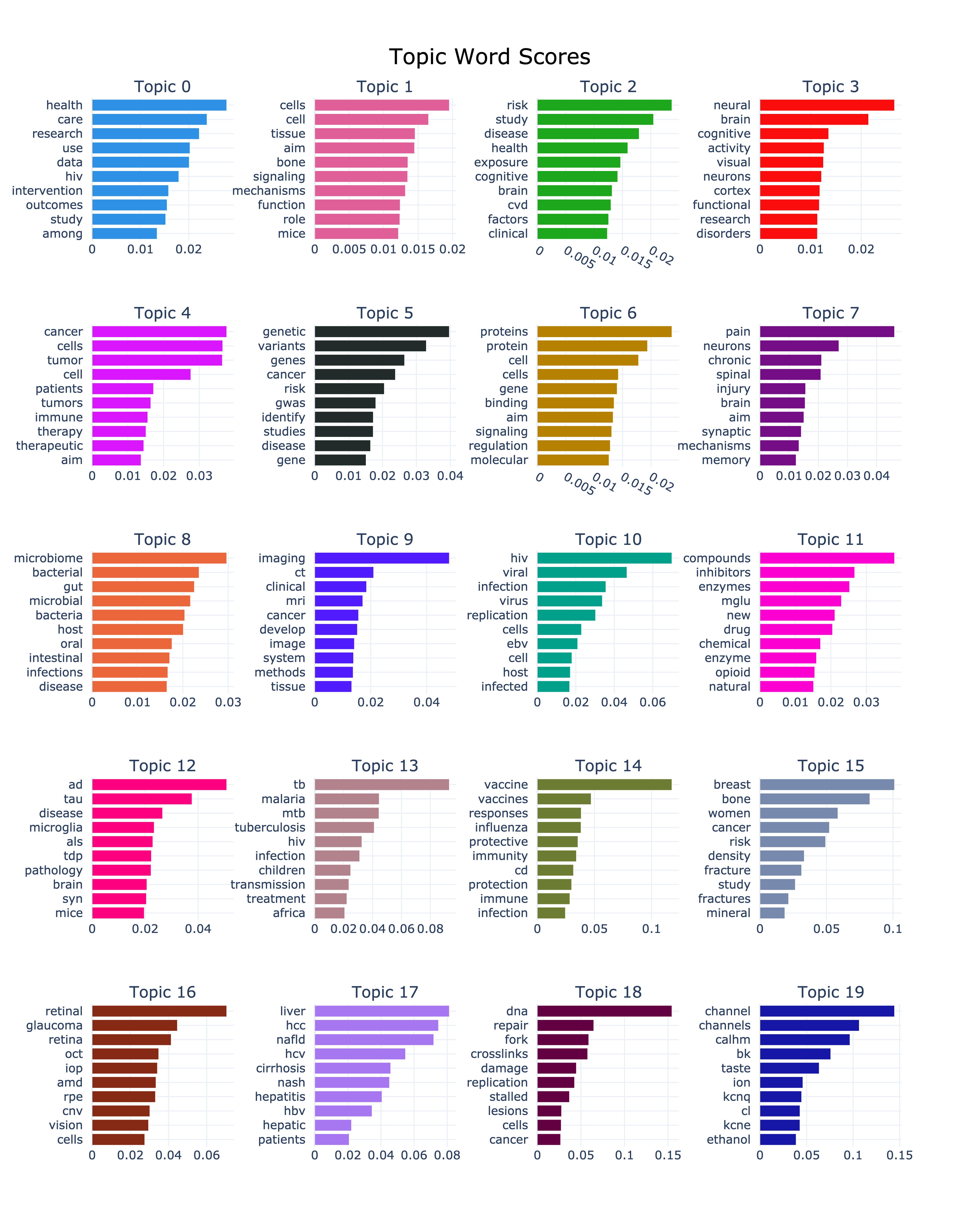}
	\caption{
Top keywords defining each BERTopic topic based on BioSentVec embedding representations, ranked by c-TF-IDF importance. Topics capture major biomedical research domains within the NIH-MPINet collaboration network, including healthcare outcomes, cardiovascular disease, neuroscience, genomics, microbiome research, virology, pharmacology, and imaging.
	}
	\label{fig:topic}
\end{figure}

\clearpage
\textbf{Topic similarity matrix.}
Figure~S2 shows the pairwise cosine similarity between topics based on BioSentVec embedding representations. 
Higher similarity values indicate that two topics share more similar semantic content in titles and abstracts. 
Several related themes show relatively higher similarity, such as molecular and cellular biology topics (Topics 1 and 6), neuroscience and neurology topics (Topics 3 and 7), and infectious disease topics (Topics 10 and 13).

\begin{figure}[htbp]
	\centering
	\includegraphics[width=0.8\textwidth]{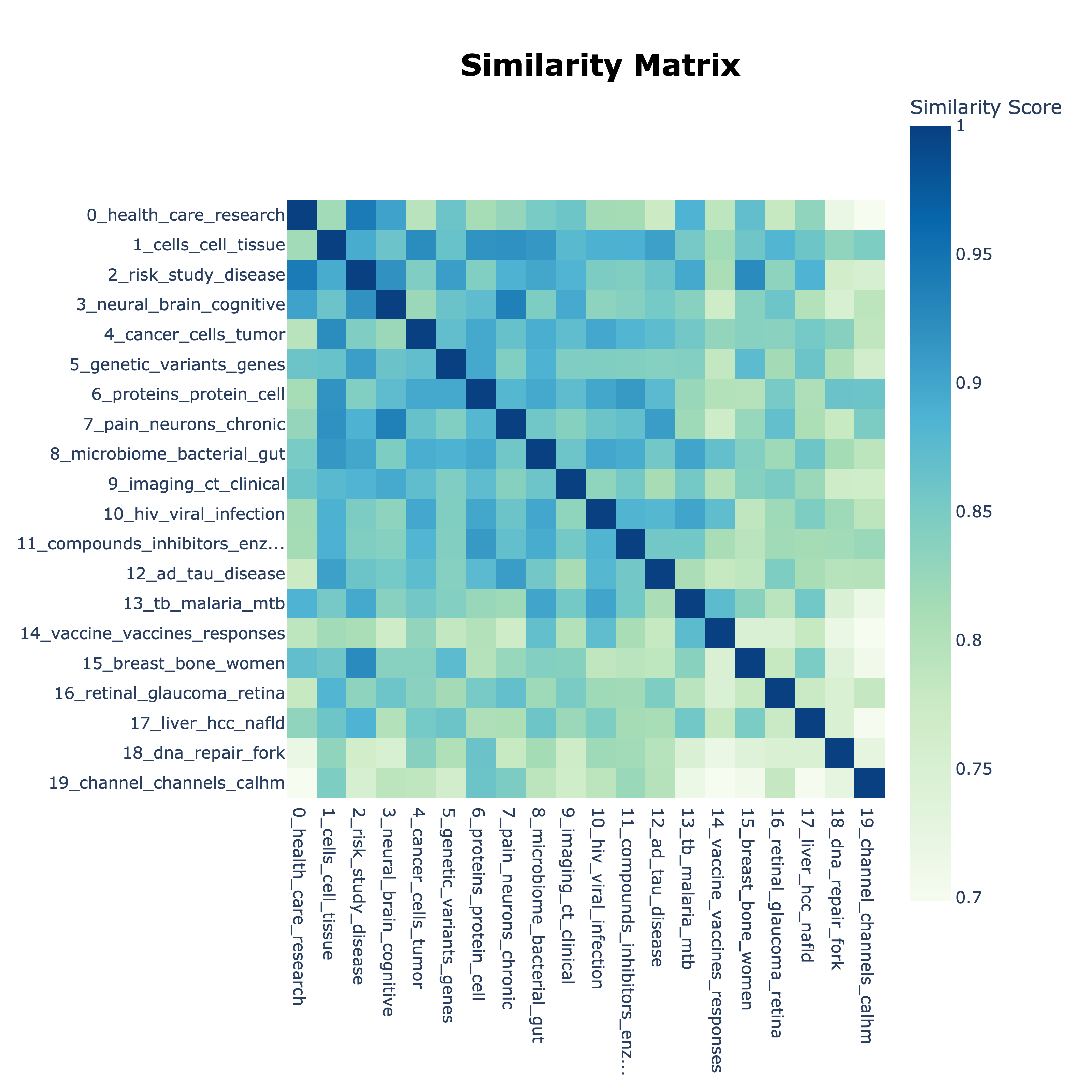}
	\caption{
Pairwise cosine similarity between BERTopic-derived topics based on BioSentVec embedding representations. Darker colors indicate stronger semantic similarity.
}
	\label{fig:similarity_matrix}
\end{figure}

\clearpage
\textbf{Dynamic BERTopic model results.}
Figure~S3 shows the normalized frequency of Topics 0–9 from 2008 to 2023, and Figure~S4 shows the normalized frequency of Topics 10–19 from 2008 to 2023. Each line represents the proportion of projects assigned to a given topic within each year, allowing comparison of relative changes over time. There were very few multi-PI projects available in 2006–2007 (3 and 23 projects, respectively). Because BERTopic requires a sufficient number of documents to identify stable clusters, most documents from these early years were classified as outliers (topic -1) and therefore do not appear in the temporal topic plots. In 2008, the number of projects increased substantially (356 projects), allowing several topics (e.g., Topic 1 and Topic 7) to be identified.

Some topics remain relatively stable across years, while others show gradual increases or decreases in prevalence. For example, Healthcare and Outcomes Research (Topic 0) and Alzheimer’s Disease and Neurodegenerative Disease (Topic 12) show a gradual increase over time, while molecular and cellular biology related topics (e.g., Topic 1 and Topic 6) appear to decline slightly after earlier peaks. 


\begin{figure}[htbp]
	\centering

	\begin{subfigure}[b]{\textwidth}
		\centering
		\includegraphics[width=\textwidth]{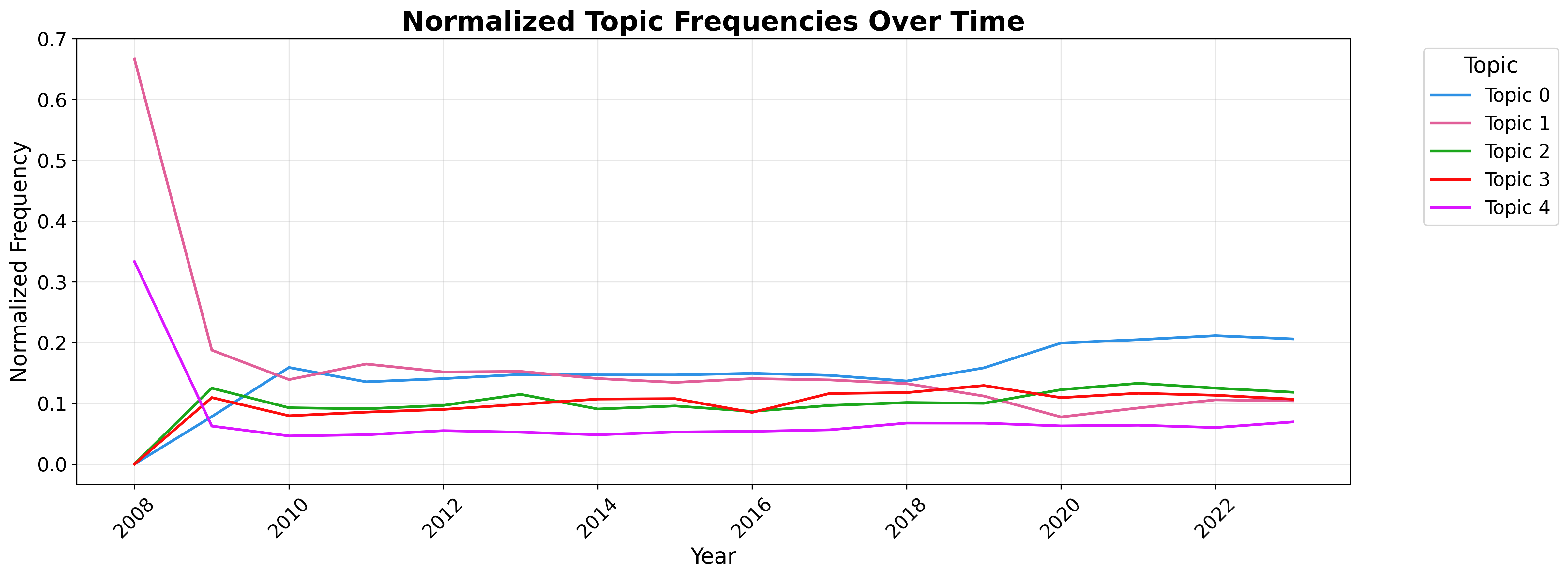}
		\label{fig:time1}
	\end{subfigure}

	\vspace{-0.5em}

	\begin{subfigure}[b]{\textwidth}
		\centering
		\includegraphics[width=\textwidth]{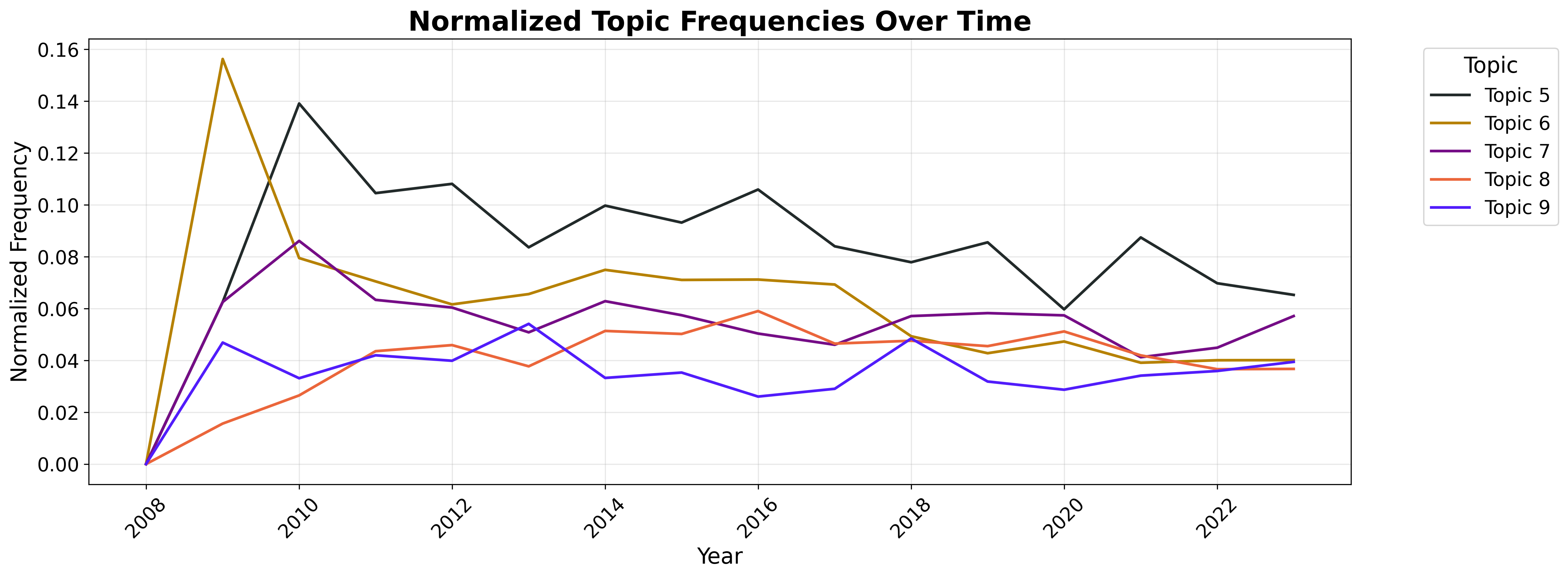}
		\label{fig:time2}
	\end{subfigure}

\caption{
Temporal evolution of BERTopic topic prevalence for Topics 0-9 from 2008–2023. Lines represent normalized topic frequencies within each year, illustrating changes in the relative prominence of major biomedical research themes over time.
}
	\label{fig:time_visualization_part1}
\end{figure}

\vspace{1em}

\begin{figure}[htbp]
	\centering

	\begin{subfigure}[b]{\textwidth}
		\centering
		\includegraphics[width=\textwidth]{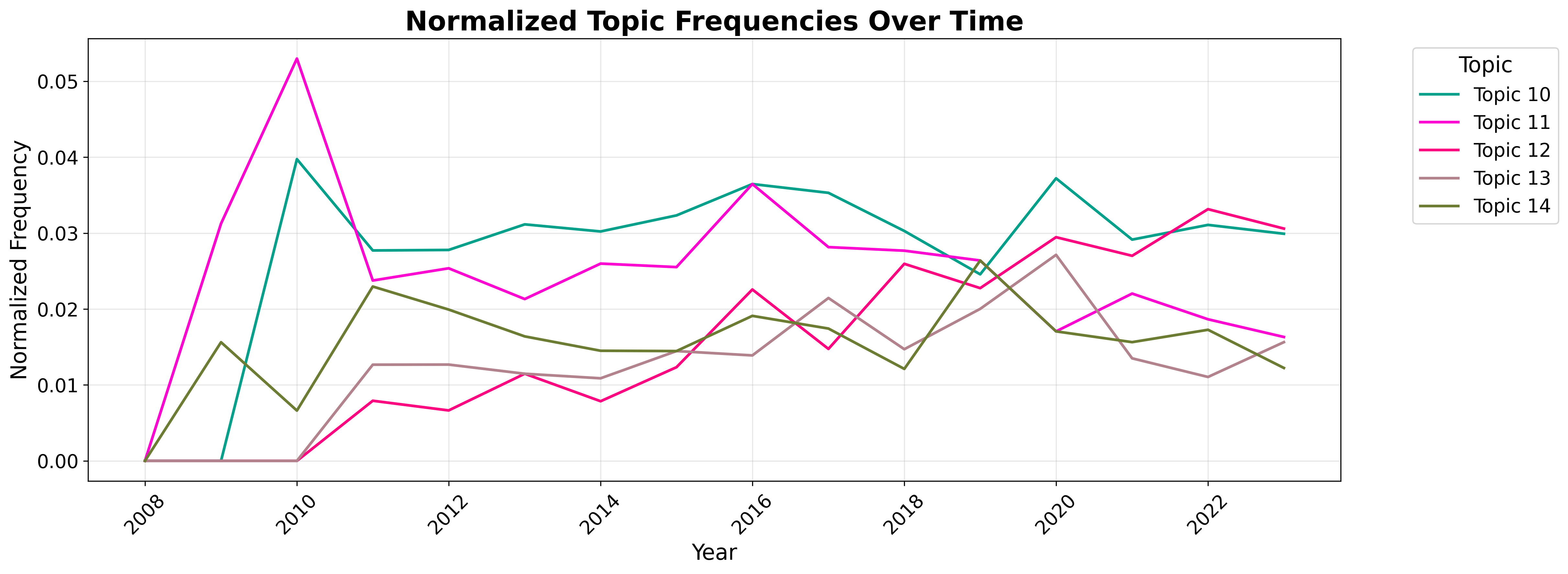}
		\label{fig:time3}
	\end{subfigure}

	\vspace{-0.5em}

	\begin{subfigure}[b]{\textwidth}
		\centering
		\includegraphics[width=\textwidth]{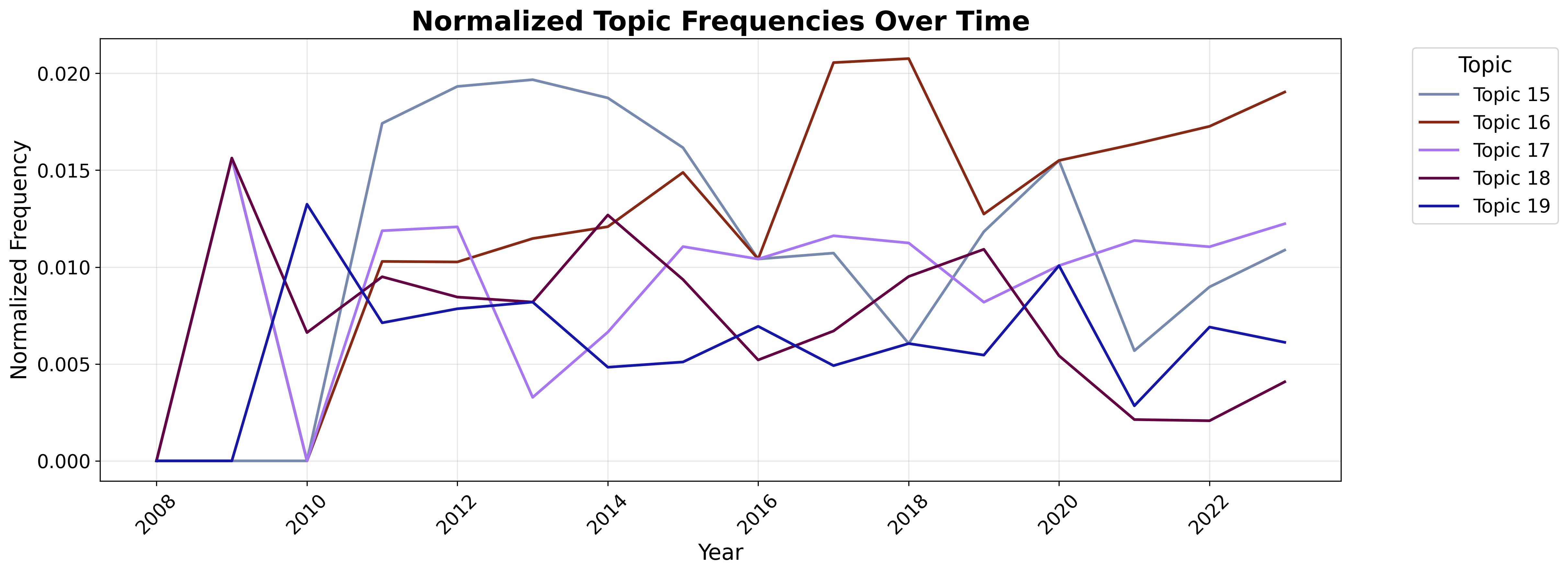}
		\label{fig:time4}
	\end{subfigure}

\caption{
Temporal evolution of BERTopic topic prevalence for Topics 10-19 from 2008–2023. Lines represent normalized topic frequencies within each year, illustrating changes in the relative prominence of major biomedical research themes over time.
}
	\label{fig:time_visualization_part2}
\end{figure}

\end{document}